\documentclass[11pt]{article}
\usepackage[final]{acl}
\usepackage{times}
\usepackage{latexsym}
\usepackage[T1]{fontenc}
\usepackage[utf8]{inputenc}
\usepackage{microtype}
\usepackage{inconsolata}
\usepackage{amsmath,amssymb,booktabs,graphicx,tabularx,xurl}
\usepackage{tikz}
\usepackage{placeins}
\usetikzlibrary{positioning,arrows.meta}

\newcommand{\appref}[1]{Appendix~\ref{#1}}

\newif\ifRDComplete
\RDCompletetrue
\newif\ifRDLatencyReady
\RDLatencyReadytrue
\newif\ifRDLambdaReady
\RDLambdaReadytrue

\newcommand{\RDResultScope}{mean $\pm$ sample SD over three seeds}
\newcommand{\RDDefaultLambda}{0.5}
\newcommand{\RDWeightAccuracyDelta}{0.30}
\newcommand{\RDWeightFlipDelta}{0.34}

\newcommand{\RDWeightSKLReductionMin}{40.0}
\newcommand{\RDWeightSKLReductionMax}{60.5}
\newcommand{\RDMMAUSKLReduction}{43.8}
\newcommand{\RDMMSUSKLReduction}{48.9}
\newcommand{\RDMMARSKLReduction}{60.5}

\newcommand{\RDSeedSentence}{All three replications are complete.}
\newcommand{\RDInterimScope}{}
\newcommand{\RDMMAU}{68.88}
\newcommand{\RDMMAR}{55.33}
\newcommand{\RDMMSU}{61.70}
\newcommand{\RDSLURP}{95.14}
\newcommand{\RDMINDS}{88.59}
\newcommand{\RDScene}{58.83}

\newcommand{\RDNSynth}{67.90}
\newcommand{\RDVADF}{0.935}
\newcommand{\RDOrigAUC}{0.895}
\newcommand{\RDFreshAUC}{0.900}
\newcommand{\RDMMAURandomSKL}{0.0697}
\newcommand{\RDMMARRandomSKL}{0.0674}
\newcommand{\RDLatency}{62.4}
\newcommand{\RDLayaLatency}{1206.9}

\title{AudioJev: Direct Audio Decisions\\with Order-Calibrated Probabilities}

\author{
  \textbf{Sihan Lv\textsuperscript{${1}$}},
  \textbf{Zhen Li\textsuperscript{${1,4}$}},
  \textbf{Zhiqi Cao\textsuperscript{${2}$}},
  \textbf{Jinshan Zhang\textsuperscript{${1,2}$}},
  \textbf{Ying Li\textsuperscript{${1,2,5}$}},
  \textbf{Meng Xi\textsuperscript{$1,2,3,5,$}\thanks{{Meng Xi is the corresponding author.}}},
  \textbf{Jianwei Yin\textsuperscript{${1,2,3,5}$}}
\\
  \small\textsuperscript{1} Zhejiang University, Hangzhou, China\\
  \small\textsuperscript{2} Innovation and Management Center of the School of Software (Ningbo), Zhejiang University, Ningbo, China\\
  \small\textsuperscript{3} Binjiang Institute of Zhejiang University, Hangzhou, China\\
  \small\textsuperscript{4} China Academy of Space Technology, Beijing, China\\
  \small\textsuperscript{5} Zhejiang Key Laboratory of Digital-Intelligence Service Technology, Hangzhou, China\\
  \small\texttt{\{shlv,lizhen\_cast,zhangjinshan,cnliying,ximeng\}@zju.edu.cn}\\
  \small\texttt{caozhiqi@zju-bj.com,zjuyjw@cs.zju.edu.cn}
}

\begin{document}
\maketitle
\begin{abstract}
Audio decisions often depend on evidence that a transcript does not preserve, while their probability estimates can depend on how answer options are ordered. AudioJev maps a waveform, question and supplied alternatives directly to a candidate distribution through one shared full-parameter model. We define \emph{order calibration} as preserving an answer's probability under meaning-preserving option permutations. Random-derangement SKL training pairs each question with a reordered view in which every alternative changes position, supervises both answers, and aligns the two distributions before applying a symmetric KL penalty. Inference retains a single candidate-scoring forward, with no calibration head or order ensemble.
\ifRDComplete
Across three training seeds, AudioJev reaches \RDMMAU\%/\RDMMAR\% mean accuracy on complete MMAU/MMAR and reduces random-order SKL by \RDMMAUSKLReduction\%/\RDMMARSKLReduction\% relative to the single-view removal ablation.
\fi
The same model handles intent, environmental sound, note properties, speech activity and conversational transitions. Paired removal ablations and multi-order evaluation measure predictive accuracy and probability stability together, establishing a direct audio interface whose calibration objective acts on candidate meaning rather than presentation position. Inference code and model weights are available at \url{https://github.com/SihanLv/AudioJev-Inference} and \url{https://huggingface.co/shlv/AudioJev}.
\end{abstract}

\section{Introduction}
\label{sec:intro}
An audio application often needs a decision rather than a transcript: which intent to route, which sound is present, whether speech is active, or who will speak next. A shared classifier can express these tasks through a question and a small set of supplied answers. For software to use its probabilities, however, the same answer should retain similar support when the alternatives are displayed in a different order.

Two representation choices matter. First, transcribe-then-classify systems restrict the classifier to a lexical intermediate, even when a decision depends on environmental sound, musical structure or conversational timing \citep{whisper,pengi,qwen25}. Second, a categorical answer set must be serialized for a language model. Position labels and sequence context can then influence which answer receives probability mass \citep{optionorder,pride}. An unchanged audio question can consequently yield different probabilities after a purely presentational edit.

AudioJev combines direct waveform classification with \emph{order-calibrated candidate probabilities}. A waveform, question and caller-supplied answer descriptions produce a distribution over exactly those answers, extending the typed-decision interface of Jev\footnote{Jev: \href{https://docs.typesafe.ai/introduction}{https://docs.typesafe.ai/introduction}} and Laya\footnote{Laya: \href{https://github.com/NandhaKishorM/laya}{https://github.com/NandhaKishorM/laya}} to audio. Candidate-only supervision keeps training and inference on the same support. The model handles semantic, acoustic, musical and conversational decisions through shared parameters rather than task-specific heads.

We train order calibration through \emph{random-derangement SKL} (RD-SKL). Each original question has a second view with identical audio, question and candidate contents, but no answer remains in its previous position. Both views receive label supervision. Their candidate distributions are restored to the same answer identities and constrained by average symmetric KL. This makes presentation consistency part of learning while retaining one model forward at inference. The mechanism builds on distributional consistency regularization \citep{rdrop}; its intervention is the supplied answer ordering, and its alignment follows candidate identity.

Our evaluation separates classification performance from order sensitivity. It includes the complete 10,000-question MMAU and 1,000-question MMAR suites, multilingual intent and speech understanding, acoustic classification, held-out sound questions, note properties and causal conversational histories. The final model reaches \RDMMAU\%/\RDMMAR\% on MMAU/MMAR under the result scope specified in Section~\ref{sec:setup}. Frozen direct-audio models and ASR--text cascades measure system-level capability; the original single-view recipe appears only as a removal ablation of the paired derangement component.

Our contributions are:
\begin{itemize}
  \item a shared full-parameter audio classifier with a candidate-conditioned probabilistic interface across semantic, acoustic and conversational tasks;
  \item an explicit order-calibration objective that couples supervised random-derangement pairs through identity-aligned symmetric KL;
  \item complete-benchmark accuracy and multi-order probability diagnostics, at fixed final training steps with paired training-recipe ablations.
\end{itemize}

\section{Related Work}
\label{sec:related}
\paragraph{Typed decisions and audio understanding.}
Jev, Laya and Jeff\footnote{Jeff: \href{https://github.com/Gestalt-Lab/jeff} {https://github.com/Gestalt-Lab/jeff}.} expose typed decisions to software. In audio, CLAP uses audio--text similarity \citep{clap,laionclap}, Pengi unifies tasks through language \citep{pengi}, and Qwen2.5-Omni supplies our waveform-language backbone \citep{qwen25}. AudioJev adapts this representation to distributions over caller-supplied answers across task families.

\paragraph{Option-order sensitivity.}
Choice ordering changes language-model decisions and probabilities \citep{optionorder,pride}. PriDe estimates option-ID priors for inference-time debiasing \citep{pride}. Our approach instead supervises paired presentations during full-parameter audio adaptation and evaluates the identity-aligned distribution, including changes that do not flip the winning answer.

\paragraph{Consistency and calibration objectives.}
R-Drop penalizes bidirectional KL between dropout-induced views \citep{rdrop}. RD-SKL uses a related consistency penalty with a semantic transformation: all supplied options change positions and outputs are inverse-bound before comparison. Conventional probability calibration relates confidence to empirical correctness \citep{guo2017calibration,ordercal}. Our operational target is \emph{order calibration}; predictive accuracy and distribution sharpness are measured alongside it. Symmetric KL itself is an established regularizer, while the experimental question is how its derangement-paired use shapes a shared audio decision interface.

\paragraph{Audio and conversational evaluation.}
SLURP and MINDS cover spoken intent \citep{slurp,minds}; MMSU, MMAU and MMAR extend evaluation to diverse speech, sound and music questions \citep{mmsu,mmau,mmar}. Silero VAD\footnote{Silero VAD: \href{https://github.com/snakers4/silero-vad} {https://github.com/snakers4/silero-vad}} and Smart Turn\footnote{Smart Turn: \href{https://github.com/pipecat-ai/smart-turn} {https://github.com/pipecat-ai/smart-turn}.} provide activity and completion references. and Observed next-speaker change remains a distinct target from completion, endpoint discovery or ideal reply timing.

\section{AudioJev}
\label{sec:method}
\begin{figure*}[t]
\centering
\includegraphics[width=\textwidth]{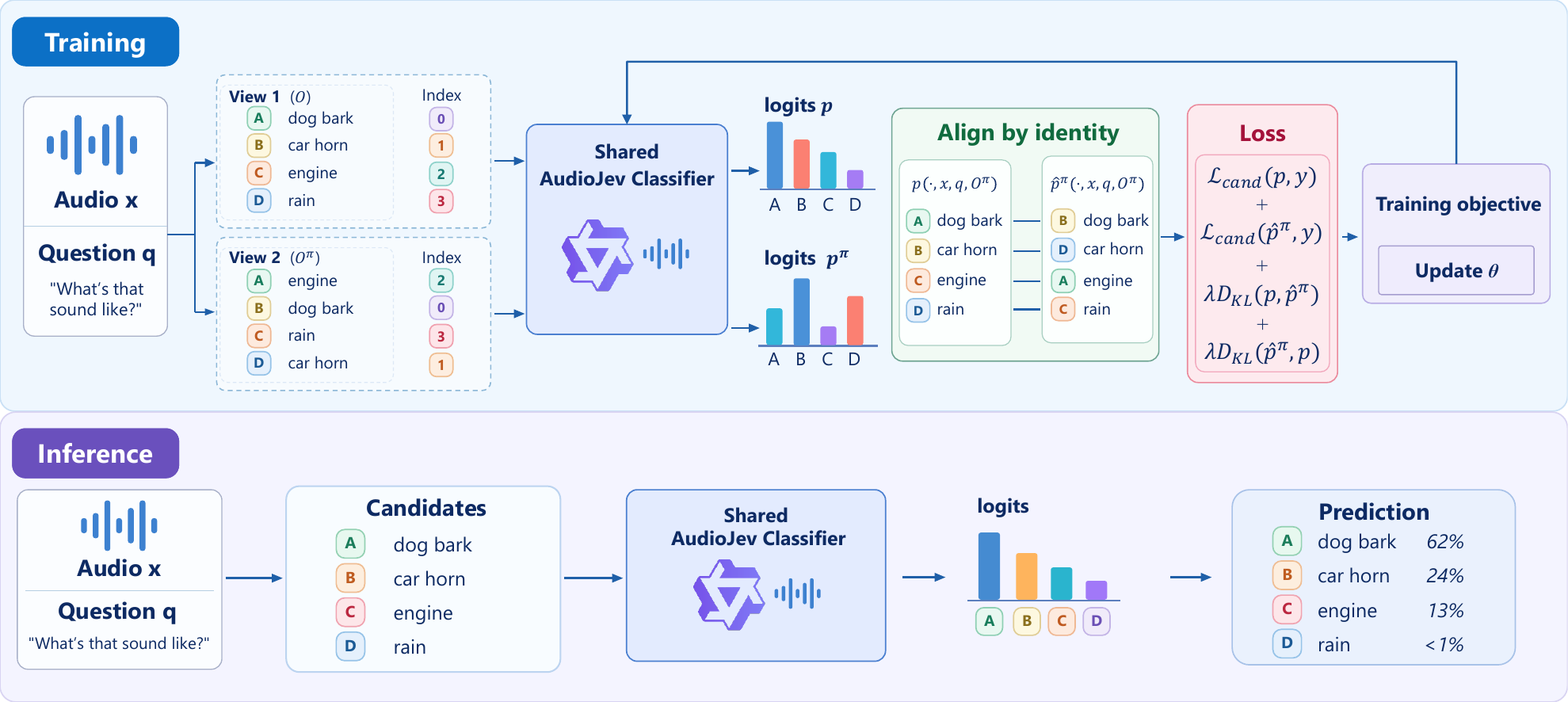}
\caption{\textbf{AudioJev overview.} A waveform, question and caller-supplied alternatives define a decision for one shared audio classifier. \textbf{Training (top):} the original and randomly deranged answer lists receive candidate-only supervision, with identity-aligned symmetric KL encouraging consistent probabilities across presentations. \textbf{Inference (bottom):} one forward pass scores the supplied alternatives and returns their probability distribution, supporting different audio tasks through the same interface.}
\label{fig:interface}
\end{figure*}

Figure~\ref{fig:interface} introduces AudioJev's training and inference interface. A shared model maps a waveform, question and supplied answer list to candidate probabilities. Paired training teaches those probabilities to follow answer meaning across order changes; inference returns them in the caller's order with one forward pass.

\subsection{Candidate-Conditioned Decisions}
Let $x$ be audio evidence, $q$ a question, and $O=(o_1,\ldots,o_K)$ the supplied answer descriptions. The output estimates $p_\theta(y\mid x,q,O)$ for one of these $K$ alternatives. Position labels come from \texttt{0--9,A--Z}; each is a distinct single token. Labels encode positions, not global class identities. The prompt requests one label, and inference reads its next-token logits without generating an explanation or an answer sequence.

For label logit $s_i$, the candidate-only readout is
\begin{equation}
 p_\theta(i\mid x,q,O)=\frac{\exp s_i}{\sum_{j=1}^{K}\exp s_j}.
 \label{eq:readout}
\end{equation}
Support is exactly the supplied list, with no implicit other class. Boolean judgments use false/true alternatives (the Noul type); ordinary choices use task-defined descriptions. This closed-set interface assumes that an appropriate answer is among the supplied candidates.

\subsection{Order Calibration and Semantic Alignment}
\label{sec:calibrationmethod}
\label{sec:canonical}
We define \emph{order calibration} as consistency of a candidate's probability under meaning-preserving permutations of the answer list. Let $O^\pi=(o_{\pi(1)},\ldots,o_{\pi(K)})$ and $p^\pi=p_\theta(\cdot\mid x,q,O^\pi)$. Its aligned distribution is
\begin{equation}
 \widetilde p^{\pi}_i=p^{\pi}_{\pi^{-1}(i)}.
 \label{eq:alignment}
\end{equation}
The target property is $p_\theta(\cdot\mid x,q,O)=\widetilde p^\pi$: probability follows answer meaning rather than its display position. Comparing equal label positions without this inverse mapping would compare different answers.

The main interface directly consumes the caller's order. Thus measured agreement reflects the trained classifier. A content-canonical wrapper is retained as a separate inference diagnostic; its deterministic sorting is not used to establish learned order calibration. The invariance domain consists of independent answer meanings. Position-relative descriptions retain their original semantics and are excluded from permutation-bias comparisons, while remaining in full accuracy denominators.

\subsection{Random-Derangement SKL Training}
\label{sec:rdskl}
A derangement moves every candidate:
\begin{equation}
 \mathcal D_K=\{\pi\in S_K:\ \pi(i)\ne i\ \forall i\}.
\end{equation}
For each training record we draw $\pi$ by shuffling until it has no fixed points. The draw is determined by the record and training seed and reused for repeated exposures. Both views use the same waveform and question; only the candidate order and its position-label binding change. In binary questions the second view is the unique swap.

Writing $p=p_\theta(\cdot\mid x,q,O)$ and $r=\widetilde p^\pi$, we use
\begin{align}
 D_{\mathrm{SKL}}(p,r)
 &=\tfrac12\bigl[D_{\mathrm{KL}}(p\Vert r)+D_{\mathrm{KL}}(r\Vert p)\bigr],
 \label{eq:skl}\\
 \mathcal L_{\mathrm{RD}}
 &=\tfrac12\bigl[\mathcal L_{\mathrm{cand}}(p,y)
                  +\mathcal L_{\mathrm{cand}}(r,y)\bigr]\notag\\
 &\quad+\lambda D_{\mathrm{SKL}}(p,r),
 \label{eq:rdloss}
\end{align}
where 
\begin{equation}
 \begin{aligned}
 \mathcal L_{\mathrm{cand}}(p,y)&=-\log p_y\\
 &=-s_y+\log\sum_{j=1}^{K}\exp s_j.
 \end{aligned}
 \label{eq:loss}
\end{equation}
Both branches receive gradients. Label supervision rewards discrimination among the supplied answers; SKL couples their probabilities across two presentations. Random derangements ensure that every candidate experiences a position change in the paired example.

RD-SKL is supervised, differentiable consistency training. It requires no sampled decision reward, separate policy head or inference-time probability map. The symmetric penalty has the same general form as established consistency regularization \citep{rdrop}; here the two views differ in option presentation and require semantic alignment before comparison.

\section{Experimental Setup}
\label{sec:setup}
\paragraph{Tasks and data.}
We evaluate semantic, acoustic, musical and conversational decisions through the same candidate interface. Three training stages use 4,096, 4,096 and 8,192 original questions, with 2--36 candidates. The final mixture covers intent, events, scenes, speech activity, speaker transitions, sound questions and note properties. Evaluation includes task-specific held-out sets, 672 multilingual MINDS recordings, 3,931 short-audio MMSU questions, complete MMAU (10,000 questions) and MMAR (1,000 questions). Benchmark answers never enter training or model selection. Dataset composition, source-disjoint partitions and causal audio windows are specified in \appref{app:data}; official scoring and the source-overlap audit appear in \appref{app:general}.

\paragraph{Training and reporting.}
We update all active audio-classification parameters of Qwen2.5-Omni 3B with Eq.~\ref{eq:rdloss}, using $\lambda=\RDDefaultLambda$, learning rate $10^{-6}$ and four original questions/eight views per update. The stages run for 1,024, 1,024 and 2,048 updates; every reported checkpoint is the fixed final step. Seeds 20261001--20261003 vary training order and derangements on identical data partitions. AudioJev results are \RDResultScope{}, with one model per prediction. The weight study compares the default and doubled weights across all three seeds. Stage schedules and implementation details are in \appref{app:generaltraining} and \appref{app:implementation}.

\paragraph{Comparisons.}
Frozen direct-audio Omni 3B/7B, CLAP, Silero VAD and Smart Turn provide backbone and specialist references. Transcript pipelines share Whisper large-v3 outputs and use Laya, Omni-text 3B/7B, FLAN-T5-XL or Qwen3.5-4B as readers (\appref{app:cascades}). The single-view CE removal ablation matches original examples, seeds, update budgets and final steps; it removes both the second view and SKL. Pairing controls are in \appref{app:rdcontrols}. Runtime measures batch-one waveform-to-probabilities on a fixed five-benchmark panel, including fresh transcription for cascades (\appref{app:cost}).

\paragraph{Metrics.}
Accuracy uses the original supplied order. Activity uses macro F1; next-speaker prediction additionally uses balanced accuracy (BA) and AUROC. Order calibration compares semantically aligned distributions under fixed reorderings, reporting SKL, total variation (TV) and semantic-answer flips. Independent random reorderings supply the main comparison; the full permutation panel and canonical-wrapper diagnostic are defined in \appref{app:ordercalibration}. Duplicate descriptions are merged for semantic comparisons. Position-relative questions remain in full accuracy but are excluded from order-bias metrics, leaving 9,968 MMAU and 991 MMAR questions. Order calibration measures presentation consistency; empirical confidence reliability is a separate property.

\section{Comparative Evaluation}
\label{sec:results}
The comparisons examine the three requirements of a shared audio decision interface: access to waveform evidence, coverage across tasks, and probabilities that remain stable when candidates are reordered. AudioJev uses $\lambda=\RDDefaultLambda$ and reports \RDResultScope{}; fixed external references are single-run results. Bold marks the best displayed mean and underlining the second distinct result in Tables~\ref{tab:transfer} and~\ref{tab:main}.

\begin{table*}[t]
\centering\small
\setlength{\tabcolsep}{3pt}
\begin{tabular*}{\textwidth}{@{\extracolsep{\fill}}lrrrrrrrr@{}}
\toprule
Model & SLURP@8 & MINDS & MMSU & MMAU & MMAR & Clotho-AQA & NSynth & Latency (ms)$\downarrow$ \\
\midrule
\multicolumn{9}{l}{\emph{Direct audio models}} \\
\textbf{AudioJev} & \raisebox{\dimexpr.5ex-.5\height\relax}{\shortstack[r]{\textbf{95.14}\\$\pm$ 0.40}} & \raisebox{\dimexpr.5ex-.5\height\relax}{\shortstack[r]{88.59\\$\pm$ 0.23}} & \raisebox{\dimexpr.5ex-.5\height\relax}{\shortstack[r]{\textbf{61.70}\\$\pm$ 0.68}} & \raisebox{\dimexpr.5ex-.5\height\relax}{\shortstack[r]{\textbf{68.88}\\$\pm$ 0.46}} & \raisebox{\dimexpr.5ex-.5\height\relax}{\shortstack[r]{\textbf{55.33}\\$\pm$ 0.84}} & \raisebox{\dimexpr.5ex-.5\height\relax}{\shortstack[r]{\textbf{90.93}\\$\pm$ 0.62}} & \raisebox{\dimexpr.5ex-.5\height\relax}{\shortstack[r]{\textbf{67.90}\\$\pm$ 0.92}} & \textbf{62.4} $\pm$ 1.2 \\
Frozen Omni 3B & 85.42 & 84.67 & 60.14 & \underline{68.64} & \underline{53.50} & \textbf{90.93} & \underline{45.70} & - \\
\midrule
\multicolumn{9}{l}{\emph{Whisper large-v3 + text classifier}} \\
Laya English & 76.56 & 54.91 & 44.90 & 38.70 & 35.60 & - & - & \underline{1206.9} $\pm$ 2.5 \\
Laya multilingual & 70.31 & 41.22 & 37.73 & 32.71 & 36.40 & - & - & 1212.8 $\pm$ 0.6 \\
Omni-text 3B & 84.64 & 88.99 & 56.96 & 55.22 & 44.30 & - & - & 1227.6 $\pm$ 1.4 \\
Omni-text 7B & \underline{89.06} & \underline{91.37} & 60.39 & 59.47 & 47.60 & - & - & 1234.0 $\pm$ 0.8 \\
FLAN-T5-XL & 79.17 & 61.16 & 50.62 & 52.80 & 42.60 & - & - & 1251.7 $\pm$ 4.2 \\
Qwen3.5-4B & 88.02 & \textbf{93.60} & \underline{60.77} & 60.74 & 52.20 & - & - & 1389.1 $\pm$ 2.0 \\
\bottomrule
\end{tabular*}

\caption{Classification accuracy (\%) and waveform-to-probabilities latency. AudioJev reports \RDResultScope{} at the fixed final step. Text pipelines use the same Whisper large-v3 transcripts; timing includes fresh ASR on the fixed five-benchmark panel. AudioJev/Laya timings are refreshed; other reader timings are same-panel references (\appref{app:cost}). Frozen Omni 3B shares AudioJev's inference architecture. Dashes denote unreported measurements.}
\label{tab:transfer}
\end{table*}

\subsection{Direct Audio versus Transcript Pipelines}
AudioJev reaches \RDMMAU\% on complete MMAU and \RDMMAR\% on MMAR (Table~\ref{tab:transfer}). These exceed the strongest displayed transcript pipeline, Whisper+Qwen3.5-4B, by 8.14 and 3.13 percentage points. Against Whisper+Laya English, the gains are 30.18 and 19.73 points. Reader choice improves the transcript route substantially, yet the direct waveform model retains an advantage on both general-audio suites. This is the setting motivating AudioJev: questions about sound and music require evidence that a lexical intermediate does not fully preserve. The comparisons measure complete systems with their respective training histories.

The same model supports spoken semantics. SLURP@8 accuracy is \RDSLURP\%, 6.08 points above the strongest displayed text reader, Omni-text 7B; MMSU accuracy is \RDMMSU\%, compared with 60.77\% for Qwen3.5-4B. MINDS adds six language varieties at \RDMINDS\% accuracy. These results connect general-audio understanding and intent routing through one candidate-conditioned classifier. Performance is also measured beyond the public MMAU subset: the hidden 9,000 questions average 68.65\%, compared with 70.90\% on the public 1,000 (\appref{app:general}), so the complete-suite result is driven primarily by hidden evaluation.

\paragraph{The cost of a waveform decision.}
RD-SKL changes training while retaining a single candidate-scoring forward at inference. On the fixed waveform-input panel, AudioJev averages \RDLatency\,ms per request, versus \RDLayaLatency\,ms for ASR+Laya English, a $19.3\times$ speedup. The cascade spends 1,177.8\,ms in transcription and 29.1\,ms in its reader (\appref{app:cost}). For requests arriving as audio, bypassing transcription therefore removes the dominant measured stage while providing the direct evidence used by the classifier.

\begin{table*}[t]
\centering\footnotesize
\setlength{\tabcolsep}{3pt}
\begin{tabular}{lrrrrrr}
\toprule
Model & Event@8 & Event@36 & Scene & VAD F1 & Orig. AUC & Fresh AUC \\
\midrule
Frozen Omni 3B & 95.25 & 84.50 & 29.50 & 0.174 & 0.584 & 0.678 \\
Frozen Omni 7B & \underline{96.00} & \underline{88.75} & 33.25 & 0.387 & 0.582 & 0.651 \\
CLAP & 94.75 & 88.50 & \underline{41.75} & 0.395 & 0.365 & 0.494 \\
Silero VAD & - & - & - & \underline{0.873} & - & - \\
Smart Turn (frozen) & - & - & - & - & \underline{0.855} & \underline{0.804} \\
AudioJev & \textbf{96.17} $\pm$ 0.88 & \textbf{91.17} $\pm$ 0.95 & \textbf{58.83} $\pm$ 1.26 & \textbf{0.935} $\pm$ 0.014 & \textbf{0.895} $\pm$ 0.002 & \textbf{0.900} $\pm$ 0.014 \\
\bottomrule
\end{tabular}

\caption{Acoustic and conversational tasks. Event and scene columns are accuracy (\%); VAD is macro F1 and transition columns are AUROC. AudioJev: \RDResultScope{}. Fixed references use their established task definitions; Smart Turn transfers completion prediction to observed next-speaker change. Dashes indicate inapplicable tasks.}
\label{tab:main}
\end{table*}

\subsection{Shared Acoustic and Semantic Capabilities}
The comparison with frozen Omni 3B shows which capabilities the shared training develops. Scene accuracy rises from 29.50\% to \RDScene\%, and note-property accuracy from 45.70\% to \RDNSynth\%---gains of 29.33 and 22.20 points (Tables~\ref{tab:main} and~\ref{tab:transfer}). SLURP@8 gains 9.72 points. Alongside these changes, complete MMAU accuracy is 68.88\% versus the backbone's 68.64\%, and Clotho-AQA remains at 90.93\%. The training thus broadens the backbone's decision capabilities while preserving its general-audio and sound-question performance in these comparisons.

The gains also extend to the specialist references. Scene accuracy exceeds CLAP's 41.75\% by 17.08 points; activity macro F1 reaches \RDVADF{}, compared with Silero's 0.873. For event classification, AudioJev obtains 96.17\% with eight candidates and 91.17\% with 36. Its margin over frozen Omni 7B grows from 0.17 to 2.42 points across those two candidate counts, showing the value of candidate-conditioned training when the supplied answer set expands. All these predictions use the same parameters as intent, sound-question and note-property classification. The task is specified by the question and candidate descriptions, without selecting a specialist head.

\subsection{Probability Stability under Reordering}
\label{sec:calibrationresults}
\begin{table*}[t]
\centering\small
\setlength{\tabcolsep}{4pt}
\begin{tabular}{lrrr}
\toprule
Task & Random SKL$\downarrow$ & Random TV$\downarrow$ & Flip (\%)$\downarrow$ \\
\midrule
SLURP (all) & 0.0387 $\pm$ 0.0064 & 0.0169 $\pm$ 0.0025 & 1.59 $\pm$ 0.37 \\
MINDS & 0.0623 $\pm$ 0.0113 & 0.0487 $\pm$ 0.0066 & 4.32 $\pm$ 1.69 \\
MMSU & 0.0805 $\pm$ 0.0123 & 0.0924 $\pm$ 0.0101 & 18.10 $\pm$ 2.29 \\
Scene & 0.0382 $\pm$ 0.0131 & 0.0602 $\pm$ 0.0086 & 6.33 $\pm$ 0.76 \\
Event/activity & 0.0157 $\pm$ 0.0046 & 0.0080 $\pm$ 0.0009 & 0.70 $\pm$ 0.19 \\
Transition & 0.0039 $\pm$ 0.0006 & 0.0114 $\pm$ 0.0055 & 0.42 $\pm$ 0.38 \\
New sessions & 0.0030 $\pm$ 0.0009 & 0.0103 $\pm$ 0.0063 & 0.33 $\pm$ 0.14 \\
Clotho-AQA & 0.0030 $\pm$ 0.0015 & 0.0055 $\pm$ 0.0002 & 0.63 $\pm$ 0.20 \\
NSynth & 0.0428 $\pm$ 0.0118 & 0.0557 $\pm$ 0.0051 & 8.72 $\pm$ 2.03 \\
MMAU full & 0.0697 $\pm$ 0.0050 & 0.0829 $\pm$ 0.0052 & 13.09 $\pm$ 1.63 \\
MMAR & 0.0674 $\pm$ 0.0108 & 0.0903 $\pm$ 0.0080 & 15.44 $\pm$ 0.56 \\
\bottomrule
\end{tabular}

\caption{Random-order calibration: \RDResultScope{}. Distributions are compared with the original ordering after semantic alignment. SKL and TV measure distribution differences; flip is the percentage of changed semantic answers. Random permutations are drawn independently of training partners. Lower is better.}
\label{tab:calibration}
\end{table*}
Table~\ref{tab:calibration} measures whether a presentational change alters the decision or its supporting probabilities. Event/activity and Clotho-AQA have semantic-flip rates of 0.70\% and 0.63\%; SLURP averages 1.59\% across its candidate counts. For binary Clotho-AQA questions, TV of 0.0055 means the selected semantic event's probability changes by 0.55 percentage points on average. These scores quantify stability in the supplied-order interface, with no sorting wrapper or order ensemble.

On complete MMAU, MMSU and MMAR, random-order SKL is 0.0697, 0.0805 and 0.0674. The MMAR value is \RDMMARSKLReduction\% below the single-view removal ablation, alongside a reduction in answer flips from 18.20\% to 15.44\% (Table~\ref{tab:rdablation}). The benefit therefore reaches both the probability distribution consumed by an application and the selected answer. Section~\ref{sec:ablations} examines the paired-training contribution across all three benchmarks and SKL weights; \appref{app:ordercalibration} tests transfer across additional permutations.

\subsection{Conversational Decisions from Causal Audio}
\label{sec:voiceagent}
Conversational decisions bring task coverage and probability stability together. The same model answers activity, intent and next-speaker questions by changing the supplied question and alternatives. Transition queries observe at most five seconds of past audio at a proposed pause, providing a decision from local evidence without waiting for a completed transcript.

\begin{table}[t]
\centering\footnotesize
\setlength{\tabcolsep}{3pt}
\begin{tabular}{lrrrr}
\toprule
Test & Acc. (\%) & Macro F1 & BA & AUROC \\
\midrule
Transition & \shortstack[r]{86.00\\$\pm$ 2.95} & \shortstack[r]{0.741\\$\pm$ 0.033} & \shortstack[r]{0.842\\$\pm$ 0.007} & \shortstack[r]{0.895\\$\pm$ 0.002} \\
New sessions & \shortstack[r]{85.67\\$\pm$ 2.90} & \shortstack[r]{0.731\\$\pm$ 0.031} & \shortstack[r]{0.823\\$\pm$ 0.022} & \shortstack[r]{0.900\\$\pm$ 0.014} \\
\bottomrule
\end{tabular}

\caption{Causal next-speaker prediction (\RDResultScope{}). BA is balanced accuracy. Macro F1 and BA describe the native argmax decision; AUROC measures the ranking of switch probabilities.}
\label{tab:timing}
\end{table}

\paragraph{Discriminating switches from holds.}
AudioJev reaches \RDOrigAUC/\RDFreshAUC{} AUROC on original/new-session tests, compared with 0.855/0.804 for Smart Turn (Table~\ref{tab:main}). The new-session gain is 0.096, and this test spans 179 previously unused sessions. Both sets contain 356 switches and 44 holds; always predicting a switch gives 89\% accuracy but 50\% BA. AudioJev instead reaches 84.2\%/82.3\% BA and 0.741/0.731 macro F1 (Table~\ref{tab:timing}), demonstrating discrimination of both outcomes under this imbalance. Its new-session AUROC of 0.900 supports transfer of the switch-probability ranking across sessions.

\paragraph{Stable probabilities for downstream decisions.}
Swapping the two alternatives changes the semantic answer on 0.42\% of original-session questions and 0.33\% of new-session questions (Table~\ref{tab:calibration}). In a binary task, TV equals the absolute change in switch probability: 0.0114/0.0103 correspond to mean changes of 1.14/1.03 percentage points. Thus the interface supplies both discriminative switch scores and resistance to candidate-order changes. A downstream voice agent can consume these probabilities alongside activity and intent outputs from the same model; the evaluated target is observed next-speaker behavior at proposed pauses.

\begin{table*}[t]
\centering\small
\setlength{\tabcolsep}{2.5pt}
\begin{tabular*}{\textwidth}{@{\extracolsep{\fill}}lrrrrrrrrr@{}}
\toprule
\multicolumn{1}{c}{} & \multicolumn{3}{c}{MMAU full} & \multicolumn{3}{c}{MMSU} & \multicolumn{3}{c}{MMAR} \\
\cmidrule(lr){2-4}\cmidrule(lr){5-7}\cmidrule(lr){8-10}
Configuration & Acc. (\%) & SKL$\downarrow$ & Flip (\%)$\downarrow$ & Acc. (\%) & SKL$\downarrow$ & Flip (\%)$\downarrow$ & Acc. (\%) & SKL$\downarrow$ & Flip (\%)$\downarrow$ \\
\midrule
w/o paired training & \raisebox{\dimexpr.5ex-.5\height\relax}{\shortstack[r]{\textbf{69.47}\\$\pm$ 0.32}} & \raisebox{\dimexpr.5ex-.5\height\relax}{\shortstack[r]{0.1239\\$\pm$ 0.0312}} & \raisebox{\dimexpr.5ex-.5\height\relax}{\shortstack[r]{14.10\\$\pm$ 1.29}} & \raisebox{\dimexpr.5ex-.5\height\relax}{\shortstack[r]{\textbf{62.55}\\$\pm$ 0.52}} & \raisebox{\dimexpr.5ex-.5\height\relax}{\shortstack[r]{0.1577\\$\pm$ 0.0402}} & \raisebox{\dimexpr.5ex-.5\height\relax}{\shortstack[r]{19.91\\$\pm$ 1.72}} & \raisebox{\dimexpr.5ex-.5\height\relax}{\shortstack[r]{\textbf{55.93}\\$\pm$ 0.42}} & \raisebox{\dimexpr.5ex-.5\height\relax}{\shortstack[r]{0.1703\\$\pm$ 0.0707}} & \raisebox{\dimexpr.5ex-.5\height\relax}{\shortstack[r]{18.20\\$\pm$ 2.94}} \\
\addlinespace[3pt]
AudioJev (RD-SKL) & \raisebox{\dimexpr.5ex-.5\height\relax}{\shortstack[r]{68.88\\$\pm$ 0.46}} & \raisebox{\dimexpr.5ex-.5\height\relax}{\shortstack[r]{\textbf{0.0697}\\$\pm$ 0.0050}} & \raisebox{\dimexpr.5ex-.5\height\relax}{\shortstack[r]{\textbf{13.09}\\$\pm$ 1.63}} & \raisebox{\dimexpr.5ex-.5\height\relax}{\shortstack[r]{61.70\\$\pm$ 0.68}} & \raisebox{\dimexpr.5ex-.5\height\relax}{\shortstack[r]{\textbf{0.0805}\\$\pm$ 0.0123}} & \raisebox{\dimexpr.5ex-.5\height\relax}{\shortstack[r]{\textbf{18.10}\\$\pm$ 2.29}} & \raisebox{\dimexpr.5ex-.5\height\relax}{\shortstack[r]{55.33\\$\pm$ 0.84}} & \raisebox{\dimexpr.5ex-.5\height\relax}{\shortstack[r]{\textbf{0.0674}\\$\pm$ 0.0108}} & \raisebox{\dimexpr.5ex-.5\height\relax}{\shortstack[r]{\textbf{15.44}\\$\pm$ 0.56}} \\
\bottomrule
\end{tabular*}

\caption{Paired derangement training versus its removal ablation (\RDResultScope{}). Original examples, seeds, updates and final steps match. Accuracy is native-order percent; SKL and flip use independent non-cyclic reordering. The ablation removes the second view and SKL together.}
\label{tab:rdablation}
\end{table*}

\ifRDLambdaReady
\begin{table*}[t]
\centering\small
\setlength{\tabcolsep}{2.5pt}
\begin{tabular*}{\textwidth}{@{\extracolsep{\fill}}lrrrrrrrrr@{}}
\toprule
\multicolumn{1}{c}{} & \multicolumn{3}{c}{MMAU full} & \multicolumn{3}{c}{MMSU} & \multicolumn{3}{c}{MMAR} \\
\cmidrule(lr){2-4}\cmidrule(lr){5-7}\cmidrule(lr){8-10}
$\lambda$ ($n$) & Acc. (\%) & SKL$\downarrow$ & Flip (\%)$\downarrow$ & Acc. (\%) & SKL$\downarrow$ & Flip (\%)$\downarrow$ & Acc. (\%) & SKL$\downarrow$ & Flip (\%)$\downarrow$ \\
\midrule
0.25 (1) & 68.97 & 0.0721 & 12.17 & 62.48 & 0.0871 & 16.87 & 55.50 & 0.0771 & 15.74 \\
\addlinespace[3pt]
0.5 (3) & \raisebox{\dimexpr.5ex-.5\height\relax}{\shortstack[r]{68.88\\$\pm$ 0.46}} & \raisebox{\dimexpr.5ex-.5\height\relax}{\shortstack[r]{0.0697\\$\pm$ 0.0050}} & \raisebox{\dimexpr.5ex-.5\height\relax}{\shortstack[r]{13.09\\$\pm$ 1.63}} & \raisebox{\dimexpr.5ex-.5\height\relax}{\shortstack[r]{61.70\\$\pm$ 0.68}} & \raisebox{\dimexpr.5ex-.5\height\relax}{\shortstack[r]{0.0805\\$\pm$ 0.0123}} & \raisebox{\dimexpr.5ex-.5\height\relax}{\shortstack[r]{18.10\\$\pm$ 2.29}} & \raisebox{\dimexpr.5ex-.5\height\relax}{\shortstack[r]{55.33\\$\pm$ 0.84}} & \raisebox{\dimexpr.5ex-.5\height\relax}{\shortstack[r]{0.0674\\$\pm$ 0.0108}} & \raisebox{\dimexpr.5ex-.5\height\relax}{\shortstack[r]{15.44\\$\pm$ 0.56}} \\
\addlinespace[3pt]
0.75 (1) & 68.95 & 0.0734 & 11.09 & 61.89 & 0.0911 & 15.47 & 54.70 & 0.0858 & 13.42 \\
\addlinespace[3pt]
1 (3) & \raisebox{\dimexpr.5ex-.5\height\relax}{\shortstack[r]{68.68\\$\pm$ 1.12}} & \raisebox{\dimexpr.5ex-.5\height\relax}{\shortstack[r]{0.0743\\$\pm$ 0.0065}} & \raisebox{\dimexpr.5ex-.5\height\relax}{\shortstack[r]{13.33\\$\pm$ 0.09}} & \raisebox{\dimexpr.5ex-.5\height\relax}{\shortstack[r]{61.99\\$\pm$ 0.79}} & \raisebox{\dimexpr.5ex-.5\height\relax}{\shortstack[r]{0.0928\\$\pm$ 0.0082}} & \raisebox{\dimexpr.5ex-.5\height\relax}{\shortstack[r]{17.89\\$\pm$ 0.45}} & \raisebox{\dimexpr.5ex-.5\height\relax}{\shortstack[r]{55.50\\$\pm$ 0.66}} & \raisebox{\dimexpr.5ex-.5\height\relax}{\shortstack[r]{0.0789\\$\pm$ 0.0129}} & \raisebox{\dimexpr.5ex-.5\height\relax}{\shortstack[r]{15.78\\$\pm$ 1.75}} \\
\addlinespace[3pt]
1.5 (1) & 68.04 & 0.0525 & 12.23 & 60.98 & 0.0571 & 17.48 & 54.00 & 0.0527 & 15.84 \\
\bottomrule
\end{tabular*}

\caption{SKL-weight sensitivity on MMAU/MMSU/MMAR. Rows with $n=3$ report mean $\pm$ sample SD; $n=1$ rows report one seed. Accuracy uses native order; SKL and flip use independent random reordering of 9,968/3,931/991 eligible questions. All runs use the fixed final checkpoint.}
\label{tab:lambda}
\end{table*}
\fi

\section{Ablations and Mechanism Analysis}
\label{sec:analysis}
\label{sec:ablations}

\subsection{Removing Paired Derangement Training}
Table~\ref{tab:rdablation} reserves the original single-view model for the removal ablation. Paired derangement training reduces random-order SKL by \RDMMAUSKLReduction\%, \RDMMSUSKLReduction\% and \RDMMARSKLReduction\% on MMAU, MMSU and MMAR, respectively, and lowers answer flips on all three. The final model retains \RDMMAU\%/\RDMMSU\%/\RDMMAR\% accuracy, linking probability consistency to useful audio decisions. These effects reflect both supervised reordering and the aligned SKL penalty.

The first-seed controls in \appref{app:rdcontrols} additionally compare cyclic pairs trained with CE, cyclic pairs with CE+SKL, and the final random-derangement recipe. The cyclic CE row tests supervised reordering, but its pairing differs from the final random-derangement row. These controls are not promoted to additional final models or task-wise alternatives.

\ifRDLambdaReady
\subsection{Effect and Robustness of the SKL Weight}
\label{sec:lambda}
Table~\ref{tab:lambda} answers two complementary questions. A sweep over five weights with one fixed training seed shows how regularization strength affects decisions and probabilities. We also train $\lambda=0.5$ and $1.0$ across three seeds to measure variation in the default and doubled weights. All models use the same data recipe, seed-dependent derangements and final checkpoint rule.
In the three-seed rows, doubling $\lambda$ changes mean accuracy by at most \RDWeightAccuracyDelta{} percentage points and mean flip rate by at most \RDWeightFlipDelta{} points on MMAU, MMSU and MMAR. Both weights reduce SKL by \RDWeightSKLReductionMin--\RDWeightSKLReductionMax\% relative to the removal ablation. Across the single-seed sweep, $\lambda=0.25$ gives the highest native-order accuracy on all three benchmarks, $\lambda=1.5$ the lowest SKL, and $\lambda=0.75$ the fewest semantic flips. This separates distributional agreement from answer stability while showing that the calibration gain persists when the default weight is doubled.
\fi

\begin{table*}[t]
\centering\small
\setlength{\tabcolsep}{3pt}
\begin{tabular*}{\textwidth}{@{\extracolsep{\fill}}llrrr@{}}
\toprule
Evaluation order & Model & MMAU SKL$\downarrow$ & MMSU SKL$\downarrow$ & MMAR SKL$\downarrow$ \\
\midrule
One-step rotation & w/o paired training & 0.1417 $\pm$ 0.0375 & 0.1826 $\pm$ 0.0549 & 0.2016 $\pm$ 0.1030 \\
 & AudioJev (RD-SKL) & \textbf{0.0750} $\pm$ 0.0057 & \textbf{0.0835} $\pm$ 0.0137 & \textbf{0.0742} $\pm$ 0.0107 \\
\addlinespace[3pt]
Half-cycle rotation & w/o paired training & 0.1509 $\pm$ 0.0359 & 0.1975 $\pm$ 0.0470 & 0.2013 $\pm$ 0.0738 \\
 & AudioJev (RD-SKL) & \textbf{0.0927} $\pm$ 0.0088 & \textbf{0.1143} $\pm$ 0.0181 & \textbf{0.0935} $\pm$ 0.0110 \\
\addlinespace[3pt]
Random non-cyclic & w/o paired training & 0.1239 $\pm$ 0.0312 & 0.1577 $\pm$ 0.0402 & 0.1703 $\pm$ 0.0707 \\
 & AudioJev (RD-SKL) & \textbf{0.0697} $\pm$ 0.0050 & \textbf{0.0805} $\pm$ 0.0123 & \textbf{0.0674} $\pm$ 0.0108 \\
\addlinespace[3pt]
Reverse & w/o paired training & 0.1671 $\pm$ 0.0409 & 0.2340 $\pm$ 0.0663 & 0.2498 $\pm$ 0.1088 \\
 & AudioJev (RD-SKL) & \textbf{0.0960} $\pm$ 0.0101 & \textbf{0.1229} $\pm$ 0.0213 & \textbf{0.1049} $\pm$ 0.0156 \\
\bottomrule
\end{tabular*}

\caption{Transfer across evaluation orders: mean $\pm$ sample SD over three paired training seeds. Each reordered distribution is compared with native order after semantic alignment, using the same 9,968/3,931/991 eligible MMAU/MMSU/MMAR questions for both models. All comparisons use fixed final checkpoints. Lower SKL is better; bold marks the lower mean within each order.}
\label{tab:ordertransfer}
\end{table*}

\subsection{What the Permutation Changes}
All deranged candidates leave their original positions, while question and audio evidence remain fixed. For $K\ge4$, random derangements can alter adjacency beyond a single cyclic operator's orbit. A finite collection of such pairs trains consistency that we evaluate under reverse, non-cyclic random and cyclic orderings.

Probability concentration also matters: flatter distributions can have smaller divergence without better decisions. We retain entropy, maximum probability, TV, flips and accuracy together, and separate the optional canonical wrapper from the learned native-order behavior. Distributional stability and predictive correctness are measured as distinct properties.

\subsection{Transfer across Evaluation Orders}
\label{sec:ordertransfer}
Each training record has one fixed derangement partner. Table~\ref{tab:ordertransfer} tests how the resulting consistency extends to four evaluation transformations: one-step rotation, half-cycle rotation, independent non-cyclic random reordering and reversal. They preserve the waveform, question and candidate meanings while changing the serialized answer list. Both models receive exactly the same evaluation views, and their outputs are aligned to the same semantic identities. Thus the comparison measures sensitivity to presentation on a common set of audio decisions.

\paragraph{Consistency across distinct rearrangements.}
RD-SKL reduces mean SKL in all 12 benchmark--order conditions. For one-step and half-cycle rotations, the reductions range from 38.6\% to 63.2\%. Under reversal, mean SKL falls by 42.6\% on MMAU, 47.5\% on MMSU and 58.0\% on MMAR. Rotations move contiguous blocks of options, whereas reversal changes their pairwise relative order. The independent random view adds non-cyclic rearrangements and yields the reductions reported in Table~\ref{tab:rdablation}. Together, these interventions show that the consistency gain transfers across several ways of presenting an unchanged answer set.

\paragraph{Agreement across seeds and distribution metrics.}
Every training seed has lower SKL for every benchmark and evaluation order, covering all 36 seed--benchmark--order comparisons. Total variation also decreases in all 36 comparisons. Across the 12 benchmark--order means, absolute TV reductions range from 0.0162 to 0.0498. SKL measures changes through probability ratios, while TV measures the amount of probability mass redistributed; their agreement supports a distribution-level benefit across both measures. The complete final-model SKL, TV and semantic-flip results remain in \appref{app:ordercalibration}.

These comparisons hold original training examples, stage budgets and final checkpoint rules fixed between the paired recipe and its removal ablation. The three seeds vary training order and the fixed per-record derangements, so the common direction extends across both training draws and evaluation transformations. Combined with the native-order accuracy in Table~\ref{tab:rdablation}, the results establish useful audio decisions with reduced sensitivity to candidate presentation. All measurements use the caller's supplied order: this transfer is expressed by the trained model in a single forward pass.

\FloatBarrier
\section{Conclusion}
AudioJev couples direct waveform classification with an explicit objective for candidate-order calibration. Random-derangement pairs expose every answer to a position change, while label supervision and identity-aligned SKL train one shared probability-producing model. The default $\lambda=0.5$ reduces order divergence across audio tasks while retaining one inference forward. Three-seed comparisons show stable mean accuracy and answer-flip rates when the weight doubles to 1.0 on MMAU, MMSU and MMAR, supporting a practical calibration objective with tolerance to regularization strength. On the fixed timing panel, one waveform question averages \RDLatency\,ms. Together, these results establish a shared audio decision interface that carries candidate meaning through training and inference across speech, sound, music and causal conversation.

\label{main-text-end}
\clearpage
\section*{Limitations}
\label{sec:limitations}
\paragraph{Decision interface.}
AudioJev targets closed-set decisions in which the caller supplies the admissible answers, the setting of typed-decision interfaces described above. Within this setting, order calibration addresses the presentation change that every serialized candidate list faces: permutation of its options. Because training pairs are aligned by candidate identity, the objective extends directly to other meaning-preserving edits, such as option paraphrases. Order calibration is complementary to confidence calibration against empirical accuracy; since RD-SKL adds no inference-time component, standard post-hoc calibrators such as temperature scaling \citep{guo2017calibration} apply on top of it unchanged.

\paragraph{Conversational deployment.}
Conversational decisions are evaluated offline from causal, past-only audio at proposed pauses, which measures the next-speaker estimate independently of any particular agent's control loop. When these probabilities drive a streaming, full-duplex agent, interruption costs and waiting thresholds are set by the deployment rather than by the classifier.

\section*{Ethical Considerations}
Audio may contain identifying information. These benchmark scores do not validate personal-attribute or identity decisions; user-affecting deployment requires application-specific validation. We use existing datasets and models, without recruiting participants or inferring actual mental states from expression labels. Dataset and model licenses vary, including research and non-commercial restrictions. Redistribution requires checking the applicable terms. The manuscript contains neither waveforms nor model weights, and AudioJev is independent of TypeSafe.

\section*{Acknowledgments}
This work is supported by the National Natural Science Foundation of China (62502427), the Science and Technology Program of Zhejiang Province (2026SDXT012), the Yongjiang Talent Introduction Program (2024A-404-G), the Major Scientific and Technological Projects of CNTC (110202401031(SZ-05)), and the Zhejiang Key Laboratory Project (2024E10001).
\bibliography{references}

@misc{qwen25,
      title={Qwen2.5-Omni Technical Report}, 
      author={Jin Xu and Zhifang Guo and Jinzheng He and Hangrui Hu and Ting He and Shuai Bai and Keqin Chen and Jialin Wang and Yang Fan and Kai Dang and Bin Zhang and Xiong Wang and Yunfei Chu and Junyang Lin},
      year={2025},
      eprint={2503.20215},
      archivePrefix={arXiv},
      primaryClass={cs.CL},
      url={https://arxiv.org/abs/2503.20215}, 
}

@inproceedings{slurp,
  title={SLURP: A spoken language understanding resource package},
  author={Bastianelli, Emanuele and Vanzo, Andrea and Swietojanski, Pawel and Rieser, Verena},
  booktitle={Proceedings of the 2020 Conference on Empirical Methods in Natural Language Processing (EMNLP)},
  pages={7252--7262},
  year={2020}
}

@inproceedings{minds,
  title={Multilingual and cross-lingual intent detection from spoken data},
  author={Gerz, Daniela and Su, Pei-Hao and Kusztos, Razvan and Mondal, Avishek and Lis, Micha{\l} and Singhal, Eshan and Mrk{\v{s}}i{\'c}, Nikola and Wen, Tsung-Hsien and Vuli{\'c}, Ivan},
  booktitle={Proceedings of the 2021 Conference on Empirical Methods in Natural Language Processing},
  pages={7468--7475},
  year={2021}
}

@inproceedings{guo2017calibration,
  title={On calibration of modern neural networks},
  author={Guo, Chuan and Pleiss, Geoff and Sun, Yu and Weinberger, Kilian Q},
  booktitle={International conference on machine learning},
  pages={1321--1330},
  year={2017},
  organization={PMLR}
}

@article{ordercal,
  title={Intra order-preserving functions for calibration of multi-class neural networks},
  author={Rahimi, Amir and Shaban, Amirreza and Cheng, Ching-An and Hartley, Richard and Boots, Byron},
  journal={Advances in neural information processing systems},
  volume={33},
  pages={13456--13467},
  year={2020}
}

@inproceedings{clap,
  title={Clap learning audio concepts from natural language supervision},
  author={Elizalde, Benjamin and Deshmukh, Soham and Al Ismail, Mahmoud and Wang, Huaming},
  booktitle={ICASSP 2023-2023 IEEE International Conference on Acoustics, Speech and Signal Processing (ICASSP)},
  pages={1--5},
  year={2023},
  organization={IEEE}
}

@article{pengi,
  title={Pengi: An audio language model for audio tasks},
  author={Deshmukh, Soham and Elizalde, Benjamin and Singh, Rita and Wang, Huaming},
  journal={Advances in Neural Information Processing Systems},
  volume={36},
  pages={18090--18108},
  year={2023}
}

@inproceedings{mmsu,
  title={Mmsu: A massive multi-task spoken language understanding and reasoning benchmark},
  author={Wang, Dingdong and Li, Junan and Wu, Jincenzi and Yang, Dongchao and Chen, Xueyuan and Zhang, Tianhua and Meng, Helen},
  booktitle={International Conference on Learning Representations},
  volume={2026},
  pages={31374--31410},
  year={2026}
}

@inproceedings{whisper,
  title={Robust speech recognition via large-scale weak supervision},
  author={Radford, Alec and Kim, Jong Wook and Xu, Tao and Brockman, Greg and McLeavey, Christine and Sutskever, Ilya},
  booktitle={International conference on machine learning},
  pages={28492--28518},
  year={2023},
  organization={PMLR}
}

@inproceedings{laionclap,
  title={Large-scale contrastive language-audio pretraining with feature fusion and keyword-to-caption augmentation},
  author={Wu, Yusong and Chen, Ke and Zhang, Tianyu and Hui, Yuchen and Berg-Kirkpatrick, Taylor and Dubnov, Shlomo},
  booktitle={ICASSP 2023-2023 IEEE International Conference on Acoustics, Speech and Signal Processing (ICASSP)},
  pages={1--5},
  year={2023},
  organization={IEEE}
}

@inproceedings{esc50,
  title={ESC: Dataset for environmental sound classification},
  author={Piczak, Karol J},
  booktitle={Proceedings of the 23rd ACM international conference on Multimedia},
  pages={1015--1018},
  year={2015}
}

@article{spokenwoz,
  title={Spokenwoz: A large-scale speech-text benchmark for spoken task-oriented dialogue agents},
  author={Si, Shuzheng and Ma, Wentao and Gao, Haoyu and Wu, Yuchuan and Lin, Ting-En and Dai, Yinpei and Li, Hangyu and Yan, Rui and Huang, Fei and Li, Yongbin},
  journal={Advances in Neural Information Processing Systems},
  volume={36},
  pages={39088--39118},
  year={2023}
}

@inproceedings{optionorder,
  title={Large language models sensitivity to the order of options in multiple-choice questions},
  author={Pezeshkpour, Pouya and Hruschka, Estevam},
  booktitle={Findings of the Association for Computational Linguistics: NAACL 2024},
  pages={2006--2017},
  year={2024}
}

@inproceedings{nsynth,
  title={Neural audio synthesis of musical notes with wavenet autoencoders},
  author={Engel, Jesse and Resnick, Cinjon and Roberts, Adam and Dieleman, Sander and Norouzi, Mohammad and Eck, Douglas and Simonyan, Karen},
  booktitle={International conference on machine learning},
  pages={1068--1077},
  year={2017},
  organization={PMLR}
}

@inproceedings{clothoaqa,
  title={Clotho-aqa: A crowdsourced dataset for audio question answering},
  author={Lipping, Samuel and Sudarsanam, Parthasaarathy and Drossos, Konstantinos and Virtanen, Tuomas},
  booktitle={2022 30th European Signal Processing Conference (EUSIPCO)},
  pages={1140--1144},
  year={2022},
  organization={IEEE}
}

@inproceedings{mmau,
  title={Mmau: A massive multi-task audio understanding and reasoning benchmark},
  author={Sakshi, Sakshi and Tyagi, Utkarsh and Kumar, Sonal and Seth, Ashish and Selvakumar, Ramaneswaran and Nieto, Oriol and Duraiswami, Ramani and Ghosh, Sreyan and Manocha, Dinesh},
  booktitle={International Conference on Learning Representations},
  volume={2025},
  pages={84929--84964},
  year={2025}
}

@article{mmar,
  title={Mmar: A challenging benchmark for deep reasoning in speech, audio, music, and their mix},
  author={Ma, Ziyang and Ma, Yinghao and Zhu, Yanqiao and Yang, Chen and Chao, Yi-Wen and Xu, Ruiyang and Chen, Wenxi and Chen, Yuanzhe and Chen, Zhuo and Cong, Jian and others},
  journal={Advances in Neural Information Processing Systems},
  volume={38},
  year={2026}
}

@article{rdrop,
  title={R-drop: Regularized dropout for neural networks},
  author={Wu, Lijun and Li, Juntao and Wang, Yue and Meng, Qi and Qin, Tao and Chen, Wei and Zhang, Min and Liu, Tie-Yan and others},
  journal={Advances in neural information processing systems},
  volume={34},
  pages={10890--10905},
  year={2021}
}

@inproceedings{pride,
  title={Large language models are not robust multiple choice selectors},
  author={Zheng, Chujie and Zhou, Hao and Meng, Fandong and Zhou, Jie and Huang, Minlie},
  booktitle={International Conference on Learning Representations},
  volume={2024},
  pages={19426--19454},
  year={2024}
}
\clearpage
\appendix
\section{Data and Three-Stage Training}
\label{app:data}
\label{app:generaltraining}
\subsection{Task Inventories and Partitions}
The final training mixture covers seven capability families (Table~\ref{tab:data}). Its 8,192 questions use 7,181 waveforms and 4,167 source groups; 576 development questions monitor optimization. An initial semantic stage and a joint acoustic stage each use 4,096 questions. Training spans 13 candidate counts between 2 and 36; event and intent questions supply the larger lists.

\begin{table}[t]
\centering\small
\setlength{\tabcolsep}{4pt}
\begin{tabular}{llr}
\toprule
Capability & Dataset & Training questions\\
\midrule
Event & ESC-50 & 1,024\\
Scene & TUT2018 & 1,024\\
Activity & Oration & 512\\
Transition & SpokenWOZ & 1,024\\
Intent & SLURP & 512\\
Sound questions & Clotho-AQA & 2,048\\
Note properties & NSynth & 2,048\\
\bottomrule
\end{tabular}
\caption{Original questions in the final general-audio stage; each has two training views. Repeated questions about a recording are not independent audio sources.}
\label{tab:data}
\end{table}

\paragraph{Acoustic and semantic tasks.}
ESC-50 uses folds 1--3 for training, fold 4 for development/calibration, and fold 5 for testing, excluding declared original-source overlap \citep{esc50}. The same 400 test clips support multiple candidate counts; this is clip classification, not temporal detection or five-fold ESC evaluation. TUT2018\footnote{TUT2018: \href{https://zenodo.org/records/1228142} {https://zenodo.org/records/1228142}.} uses recording-location-disjoint splits and 400 test clips. Oration\footnote{Oration: \href{https://huggingface.co/datasets/Oration/voice-activity-detection} {https://huggingface.co/datasets/Oration/voice-activity-detection}.} activity labels concern the final 200\,ms of one-second past-only histories; its 271 test windows come from three source recordings. SLURP uses an internal speaker-disjoint split of official training audio, with 384 test sources and candidate counts 2, 4, 8, 16 and 36 \citep{slurp}. A separate publisher-test check covers 512 recordings at 8/36 candidates. Clotho-AQA provides unanimous binary sound answers and NSynth provides note properties \citep{clothoaqa,nsynth}; their held-out suites contain 1,058 and 512 questions, respectively.

\paragraph{Conversational decisions.}
At a SpokenWOZ pause \citep{spokenwoz}, a query occurs 200\,ms after a user-word offset and observes at most five seconds of past audio. The target records whether the partner speaks next or the same user resumes within one second. Sessions do not cross partitions. Original and new-session tests contain 400 windows each, including 356 switches and 44 holds; the latter uses 179 previously unused sessions. This evaluates observed next-speaker behavior at proposed pauses, not continuous endpoint discovery.

\subsection{Stage Schedule}
All final models use the same original question inventories and development partitions. Seeds vary example order and the fixed per-record derangement. Semantic initialization uses intent questions, joint training combines acoustic tasks with intent rehearsal, and general training adds sound questions and note properties. All stages use the shared classifier and the objective in Eq.~\ref{eq:rdloss}. The stage schedule is:
\begin{center}\small
\begin{tabular}{lrrr}
\toprule
Stage & Original examples & Updates & Dev.\\
\midrule
Semantic & 4,096 & 1,024 & 256\\
Joint & 4,096 & 1,024 & 320\\
General & 8,192 & 2,048 & 576\\
\bottomrule
\end{tabular}
\end{center}
Each original example supplies two supervised views. Global batch size is four original questions/eight views. Original-example and update budgets match the removal ablation, while its single-view compute is lower. The objective and optimizer schedule are fixed across seeds; implementation details of the two-view forward are given in \appref{app:implementation}. Candidate labels are verified single tokens, and no implicit fallback category is introduced. Optimizers reset at stage boundaries, and stages continue from their fixed final step. The data order follows seed $+1000\times$ stage index, with identical original-example manifests for the final model and its removal ablation.

The activity input consists of one-second past-only histories. Speaker-change questions use at most five seconds of history at a proposed conversational pause. Local queries observe only $x_{[t-h,t]}$, and cropping precedes resampling so that the filter cannot use future samples. Activity labels concern the history's end; later annotations determine transition targets but never enter the model input. Clip-level questions retain the full permitted clip. Both views of a training pair share audio preprocessing.

\section{Training-Seed Replication}
\label{app:replicas}
The three-seed native-order scores are listed in Table~\ref{tab:seedreplicas}; each row uses the fixed final general-training step.
\begin{table}[t]
\centering\small
\setlength{\tabcolsep}{3pt}
\begin{tabular}{lrrrr}
\toprule
Seed & General step & MMAU & MMSU & MMAR \\
\midrule
01 & 2048 & 69.41 & 62.45 & 56.30 \\
02 & 2048 & 68.66 & 61.13 & 54.80 \\
03 & 2048 & 68.56 & 61.51 & 54.90 \\
\bottomrule
\end{tabular}

\caption{Final RD-SKL models: fixed final general step and native-order accuracy (\%). Seed labels abbreviate 20261001--20261003. }
\label{tab:seedreplicas}
\end{table}
\RDSeedSentence{} Reports average each benchmark's per-seed score; they do not average predictions or treat repeated seed/question pairs as independent examples. The same evaluation permutations are fixed across all seeds. Every seed uses general step 2,048 after all three stages, as does its original-recipe removal ablation. Development scores do not select these reported checkpoints.

The broad weight sweep uses one fixed training seed. The default $\lambda=0.5$ and doubled weight are each evaluated with all three seeds on the fixed benchmark inventory. Replication measures variation with training order and derangements on common test partitions.

\section{Full Benchmark Scoring and External References}
\label{app:general}
\label{app:cascades}
We evaluate 672 MINDS recordings across six language varieties \citep{minds}, 3,931 short-audio MMSU questions spanning 39 families \citep{mmsu}, all 10,000 MMAU questions, and all 1,000 MMAR questions \citep{mmau,mmar}. MMAU combines correct counts from 1,000 public questions and 9,000 hidden questions scored by the official API. MMAR uses the publisher word-set scorer. Full waveforms, original options, duplicate descriptions, padding conventions and multi-answer references are retained. Benchmark answers never enter training, checkpoint selection or calibration fitting.
The public, hidden and full MMAU scores for each seed are reported in Table~\ref{tab:mmaupartitions}.

\begin{table}[t]
\centering\small
\begin{tabular}{lrrr}
\toprule
Seed & Public (1k) & Hidden (9k) & Full (10k) \\
\midrule
01 & 71.10 & 69.22 & 69.41 \\
02 & 71.10 & 68.39 & 68.66 \\
03 & 70.50 & 68.34 & 68.56 \\
\bottomrule
\end{tabular}

\caption{MMAU partition accuracy (\%) for the final method. Full accuracy adds public and hidden correct counts before division by 10,000.}
\label{tab:mmaupartitions}
\end{table}
The hidden MMAU submissions contain every publisher ID exactly once and one selected original option description. Each of the six evaluation views has its own 9,000-question official receipt with zero missing predictions. No hidden item correctness is inferred from aggregate scores. Public-mini questions are scored with the publisher word-set rule, as are all 1,000 MMAR questions. Padding, duplicate descriptions and non-single-answer references are retained; no candidate is synthesized from a gold answer.

\section{Complete Native and Canonical Accuracy}
\label{app:binding}
Table~\ref{tab:allaccuracy} gives the complete native-versus-canonical scores, while Table~\ref{tab:heldout} reports the per-seed held-out sound and note-property results.
\begin{table}[t]
\centering\footnotesize
\setlength{\tabcolsep}{3pt}
\begin{tabular}{lrr}
\toprule
Dataset & Native & Canonical \\
\midrule
SLURP (all) & 94.62 $\pm$ 0.29 & 94.64 $\pm$ 0.49 \\
MINDS & 88.59 $\pm$ 0.23 & 88.64 $\pm$ 0.48 \\
MMSU & 61.70 $\pm$ 0.68 & 62.27 $\pm$ 1.04 \\
Scene & 58.83 $\pm$ 1.26 & 59.17 $\pm$ 3.11 \\
Scene (permuted) & 59.25 $\pm$ 2.78 & 59.17 $\pm$ 3.11 \\
Event/activity & 95.96 $\pm$ 0.56 & 95.90 $\pm$ 0.55 \\
Transition & 86.00 $\pm$ 2.95 & 85.75 $\pm$ 3.28 \\
New sessions & 85.67 $\pm$ 2.90 & 85.50 $\pm$ 2.54 \\
Long MMSU & 64.62 $\pm$ 1.33 & 64.10 $\pm$ 1.94 \\
Publisher SLURP & 91.89 $\pm$ 0.26 & 91.89 $\pm$ 0.93 \\
Clotho-AQA & 90.93 $\pm$ 0.62 & 90.93 $\pm$ 0.62 \\
NSynth & 67.90 $\pm$ 0.92 & 68.23 $\pm$ 1.39 \\
MMAU hidden & 68.65 $\pm$ 0.49 & 67.86 $\pm$ 0.61 \\
MMAU public & 70.90 $\pm$ 0.35 & 70.87 $\pm$ 1.10 \\
MMAU full & 68.88 $\pm$ 0.46 & 68.16 $\pm$ 0.65 \\
MMAR & 55.33 $\pm$ 0.84 & 56.67 $\pm$ 0.75 \\
\bottomrule
\end{tabular}

\caption{Final-model accuracy (\%): \RDResultScope{}. Native order is the main interface. Canonical order is a separate deterministic-wrapper diagnostic.}
\label{tab:allaccuracy}
\end{table}
\begin{table}[t]
\centering\small
\begin{tabular}{lrr}
\toprule
Seed & Clotho & NSynth \\
\midrule
01 & 90.36 & 68.95 \\
02 & 91.59 & 67.19 \\
03 & 90.83 & 67.58 \\
\bottomrule
\end{tabular}

\caption{Held-out sound questions and note properties for each final RD-SKL seed: accuracy (\%).}
\label{tab:heldout}
\end{table}
The canonical path sorts supported unique descriptions by their exact text and restores probabilities to caller identities. Its historical native fallback handles unsupported descriptions. It is not used in the primary order-calibration metrics, because identical sorted prompts can produce agreement independently of the learned regularizer.

Publisher SLURP samples 512 recordings from SLURP's official test partition, none of which appears in our training or development material. It complements the speaker-disjoint internal split with a check on the publisher's own partition.

\section{Pairing-Recipe Controls}
\label{app:rdcontrols}
The fixed seed01 pairing comparison is shown in Table~\ref{tab:rdcontrols}; it isolates the effects of cyclic versus random derangement pairs and the SKL term.
\begin{table*}[t]
\centering\footnotesize
\setlength{\tabcolsep}{3pt}
\begin{tabular}{lrrrrrr}
\toprule
Seed01 configuration & MMAU acc. & SKL$\downarrow$ & MMSU acc. & SKL$\downarrow$ & MMAR acc. & SKL$\downarrow$ \\
\midrule
Ablation: no paired training & 69.55 & 0.0879 & \textbf{63.14} & 0.1145 & \textbf{56.40} & 0.1055 \\
Cyclic pairs, CE & 69.87 & 0.1182 & 61.51 & 0.1534 & 56.10 & 0.1167 \\
Cyclic pairs, CE+SKL ($\lambda=1$) & 69.48 & \textbf{0.0547} & 61.56 & \textbf{0.0631} & 54.00 & \textbf{0.0509} \\
Deranged pairs, CE+SKL ($\lambda=1$) & \textbf{69.94} & 0.0667 & 61.82 & 0.0869 & 56.20 & 0.0696 \\
AudioJev: deranged pairs ($\lambda=0.5$) & 69.41 & 0.0644 & 62.45 & 0.0742 & 56.30 & 0.0548 \\
\bottomrule
\end{tabular}

\caption{Seed01 pairing controls with explicit SKL weights. Accuracy is native-order percent; SKL measures an independent non-cyclic random reordering. All two-view arms share original examples and update budgets. The single-view row is the removal ablation. The cyclic CE row is not a matched random-derangement CE control, so differences against it are recipe effects.}
\label{tab:rdcontrols}
\end{table*}
A one-step shift is a derangement with a single cycle, whereas a general random derangement may have multiple cycles. Small cyclic shifts preserve more of the list's relative ordering. The fixed comparison evaluates these pairing choices empirically; it does not assume that smaller displacement must give better optimization or arbitrary-permutation generalization. Only the random-derangement SKL recipe defines the final model.

\section{End-to-End Runtime}
\label{app:cost}
Table~\ref{tab:latencycomponents} breaks the fixed-panel runtime into transcription, reader and end-to-end components.
\begin{table*}[t]
\centering\small
\begin{tabular}{lrrrrr}
\toprule
Pipeline & ASR & Reader & Total & p50 & p95 \\
\midrule
AudioJev (RD-SKL) & \textbf{0.0} & 62.4 & \textbf{62.4} $\pm$ 1.2 & \textbf{59.6} & \textbf{84.7} \\
ASR + Laya English & 1177.8 & 29.1 & 1206.9 $\pm$ 2.5 & 557.3 & 7248.0 \\
ASR + Laya multilingual & 1188.8 & \textbf{24.1} & 1212.8 $\pm$ 0.6 & 564.1 & 7304.1 \\
\bottomrule
\end{tabular}

\caption{Fresh final-model and ASR--Laya timing on the fixed panel, in milliseconds. Total means include sample SD across three repetition means. \ifRDLatencyReady\else Pending cells are not measurements of the final model.\fi}
\label{tab:latencycomponents}
\end{table*}
The panel contains 100 distinct-recording questions, 20 each from SLURP@8, MINDS, MMSU, MMAU, and MMAR. Four independent RTX 5880 workers use batch size 1 and four CPU threads. Two warm-up requests are excluded; every assigned query is then measured three times with device synchronization.

The final model uses native-order candidate readout and the predesignated seed01 checkpoint. Its trained consistency objective adds no inference-time head. Cascades execute fresh Whisper transcription with a co-resident reader; no waveform features or transcripts are reused. Reader-only times describe applications in which text already exists. Other fixed-reader timing entries in the main comparison come from the same panel/protocol's reference runs, not a synchronized all-model speedup experiment. Each timed request performs a fresh audio encoding.

\section{Order-Calibration Definitions and Implementation}
\label{app:calibration}
\label{app:ordercalibration}
The target of order calibration is candidate-probability stability under semantically unchanged option permutations. Let $p$ and $r$ be distributions restored to the same answer identities. Besides the training penalty in Eq.~\ref{eq:skl}, evaluation reports
\begin{align}
 D_{\mathrm{JS}}(p,r)
 &=\tfrac12[D_{\mathrm{KL}}(p\Vert m)+D_{\mathrm{KL}}(r\Vert m)],\\
 m&=\tfrac12(p+r),\\
 \mathrm{TV}(p,r)&=\tfrac12\sum_i|p_i-r_i|.
\end{align}
SKL is the average of the two direct KL directions; it is not JS. All logs are natural, and the JS field reports divergence rather than its square root. Both gradients of the SKL training term are retained.

Maximum probability change is $\max_i|p_i-r_i|$. Semantic flip compares the actual selected descriptions, retaining display-order tie behavior. Top-tie and ID-level diagnostics are saved separately. Exactly duplicate descriptions are aggregated into one event before semantic distribution comparisons. This prevents probability redistribution between identical texts from being misinterpreted as a change in answer meaning. Entropy and maximum-probability diagnostics retain the original candidate distribution; semantic divergence separately merges duplicate descriptions.

\subsection{Training and Evaluation Permutations}
Training uses one fixed seed-dependent derangement per record. Each pair changes positions but not audio, question or candidate content. The CE target moves with its semantic answer. Random-order evaluation draws independently from the training partner and excludes cyclic permutations for $K>2$. For $K=2$, every non-identity permutation is the same swap; half-cycle, reverse and random views can coincide. Repeated identical views reuse a forward and do not create new independent questions.

Each question has six evaluation views: original order, one-step cyclic shift, half-cycle shift, a fixed non-cyclic random permutation, reverse order, and content-canonical order. The same permutations are used for every model and training seed. Canonical scores are a separate wrapper diagnostic. Random-order views test presentations not used as that question's training partner.

All six view predictions, alignment maps and candidate probabilities are retained. The same permutation panel is reused across final-model seeds and removal ablations. No test label enters the permutation generator, model selection rule or consistency objective.

Table~\ref{tab:ordertransfer} in Section~\ref{sec:ordertransfer} compares the final model and the removal ablation across orders on MMAU, MMSU and MMAR. Table~\ref{tab:allorderviews} supplements that comparison with TV, semantic flips, scene classification and note-property questions.

\begin{table*}[t]
\centering\small
\begin{tabular}{lrrr}
\toprule
Task/order & SKL$\downarrow$ & TV$\downarrow$ & Flip (\%)$\downarrow$ \\
\midrule
MMAU full: one-step & 0.0750 $\pm$ 0.0057 & 0.0890 $\pm$ 0.0059 & 14.04 $\pm$ 2.14 \\
MMAU full: half-cycle & 0.0927 $\pm$ 0.0088 & 0.1025 $\pm$ 0.0090 & 16.31 $\pm$ 2.99 \\
MMAU full: random & 0.0697 $\pm$ 0.0050 & 0.0829 $\pm$ 0.0052 & 13.09 $\pm$ 1.63 \\
MMAU full: reverse & 0.0960 $\pm$ 0.0101 & 0.1036 $\pm$ 0.0077 & 16.64 $\pm$ 2.88 \\
MMSU: one-step & 0.0835 $\pm$ 0.0137 & 0.0977 $\pm$ 0.0121 & 19.68 $\pm$ 2.52 \\
MMSU: half-cycle & 0.1143 $\pm$ 0.0181 & 0.1189 $\pm$ 0.0161 & 23.63 $\pm$ 3.17 \\
MMSU: random & 0.0805 $\pm$ 0.0123 & 0.0924 $\pm$ 0.0101 & 18.10 $\pm$ 2.29 \\
MMSU: reverse & 0.1229 $\pm$ 0.0213 & 0.1214 $\pm$ 0.0163 & 23.97 $\pm$ 3.48 \\
MMAR: one-step & 0.0742 $\pm$ 0.0107 & 0.0971 $\pm$ 0.0079 & 16.72 $\pm$ 2.13 \\
MMAR: half-cycle & 0.0935 $\pm$ 0.0110 & 0.1114 $\pm$ 0.0083 & 19.54 $\pm$ 2.25 \\
MMAR: random & 0.0674 $\pm$ 0.0108 & 0.0903 $\pm$ 0.0080 & 15.44 $\pm$ 0.56 \\
MMAR: reverse & 0.1049 $\pm$ 0.0156 & 0.1164 $\pm$ 0.0098 & 20.35 $\pm$ 2.91 \\
Scene: one-step & 0.0293 $\pm$ 0.0068 & 0.0544 $\pm$ 0.0088 & 5.58 $\pm$ 2.24 \\
Scene: half-cycle & 0.0526 $\pm$ 0.0334 & 0.0728 $\pm$ 0.0271 & 7.67 $\pm$ 2.75 \\
Scene: random & 0.0382 $\pm$ 0.0131 & 0.0602 $\pm$ 0.0086 & 6.33 $\pm$ 0.76 \\
Scene: reverse & 0.0448 $\pm$ 0.0187 & 0.0679 $\pm$ 0.0152 & 6.75 $\pm$ 0.66 \\
NSynth: one-step & 0.0392 $\pm$ 0.0109 & 0.0543 $\pm$ 0.0082 & 8.14 $\pm$ 2.60 \\
NSynth: half-cycle & 0.0537 $\pm$ 0.0314 & 0.0616 $\pm$ 0.0112 & 9.18 $\pm$ 2.73 \\
NSynth: random & 0.0428 $\pm$ 0.0118 & 0.0557 $\pm$ 0.0051 & 8.72 $\pm$ 2.03 \\
NSynth: reverse & 0.0628 $\pm$ 0.0247 & 0.0678 $\pm$ 0.0112 & 9.64 $\pm$ 2.85 \\
\bottomrule
\end{tabular}

\caption{Multi-order diagnostics for the final method: \RDResultScope{}. Probability differences are computed against native order after semantic alignment; flip rates are percentages.}
\label{tab:allorderviews}
\end{table*}

\subsection{Meaning-Preserving Scope}
Statements that reference option positions can change meaning after a permutation. A fixed conservative wording screen flags potentially relative references in 32 MMAU and 9 MMAR questions. Full accuracy retains denominators 10,000/1,000, while the corresponding bias denominators are 9,968/991. Candidate-count diagnostics additionally separate binary questions and $K\ge4$. Flagged questions retain the native input for the five direct-order accuracy views and are excluded from the corresponding bias denominator, with their identities and reasons recorded before inference. The separate canonical diagnostic retains the historical wrapper rule. Full benchmark accuracy never excludes these questions.

Ordinal alternatives, missing correct answers and paraphrased candidates are different interface interventions from the fixed-content permutations studied here. Finite-pair training supplies empirical consistency rather than a proof of equality over all possible lists or rephrasings.

\subsection{Optimization and Measurement}
\label{app:implementation}
The main three-seed recipe uses candidate-only CE averaged over two views plus SKL with weight $\lambda=\RDDefaultLambda$ throughout all stages. There is no reinforcement-learning estimator, reference-policy anchor or fitted monotone probability map. The active audio-classification parameters are updated jointly: audio encoder and projection, text embeddings and decoder, and output head. Unused visual and speech-synthesis branches are outside this training path.

We train on four 48\,GB RTX 5880 GPUs using FSDP. Computation uses BF16, with FP32 master parameters, Adam states and gradient reductions, AdamW decay .01 and norm clipping at 1. The inherited development-selected learning rate is $10^{-6}$, with 32 warmup updates per stage. Development measurements monitor optimization; reported checkpoints use the fixed final step, without per-task selection.

Training projects the causal state immediately before the supervised label; the attention mask excludes that label from its prediction. Inference supplies no target and returns the candidate softmax directly.

All three default-weight runs pack both views into one left-padded forward and backward pass using FSDP gradient sharding, without activation checkpointing. Checkpoints preserve optimizer and random-generator states for continuation. Checks verify target remapping, causal label exclusion, finite gradients and parameter fingerprints; fused candidate outputs are checked against single-view computation. The $\lambda=1$ comparison retains its completed runs: seed 01 used cached-logit replay, seed 02 switched from replay to fused computation at semantic step 768, and seed 03 used fused computation throughout. Both implementations differentiate the paired objective.

Accuracy and probability stability are reported together. A uniform predictor can achieve zero order divergence, and a model can preserve its winning answer while changing its probability substantially. Entropy, maximum probability, TV, SKL and answer flips therefore complement one another; none alone establishes an application-level correctness guarantee. Conventional reliability metrics address a separate confidence-versus- correctness question and are not substituted for the order-calibration target.

\end{document}